\documentclass[conference]{IEEEtran}

\usepackage{graphicx}
\usepackage{pgfplots,amsmath,amssymb}
\usepackage{amsthm}
\usepackage{mathtools}
\usetikzlibrary{shapes,arrows,fit,calc,positioning,automata}
\usepackage{tikz-3dplot}

\newcommand{\mti}{\ensuremath{{m}_\text{r}}}
\newcommand{\mtq}{\ensuremath{{m}_\text{i}}}
\newcommand{\rbicm}{\ensuremath{ R_{\text{BMD}} } }

\begin{document}

\title{Polar Codes with Integrated Probabilistic Shaping for 5G New Radio}

\author{\IEEEauthorblockN{Onurcan \.{I}\c{s}can, Wen Xu}
	\IEEEauthorblockA{Huawei Technologies D\"usseldorf GmbH,  German Research Center\\
		Riesstr. 25 80992 Munich, Germany\\
		Email: \{Onurcan.Iscan, Wen.Dr.Xu\}@Huawei.com }}
\maketitle

\begin{abstract} 
A modification to 5G New Radio (NR) polar code is proposed, which improves the error correction performance with higher order modulation through probabilistic shaping. The presented scheme mainly re-uses existing hardware at the transmitter, and modifications at the receiver are small. Simulation results show that the presented approach can improve the performance by up to 1dB for 256-QAM on AWGN channels\footnote{This work is accepted for publication at IEEE 88th Vehicular Technology Conference (VTC 2018-Fall). Copyright IEEE 2018.}.
\end{abstract}

\section{Introduction}
In order to approach the information theoretical limits, the transmitted symbols need to have a capacity achieving probability distribution. For example, Gaussian distributed symbols are required on AWGN channels and uniform signaling causes the so-called \textit{shaping loss} of up to 1.53dB \cite{kschischang1993optimal}. The existing 2G, 3G and 4G systems all use uniformly distributed transmit symbols, and therefore suffer from the shaping loss. 
Recently, 3GPP completed the first release of the fifth generation (5G) new radio (NR) standard. This new release uses bit-interleaved coded modulation (BICM) similar to 4G, but replaces the existing error correction schemes with LDPC and polar codes \cite{Arikan09} for improved performances \cite{chan_code5G}. However, similar to the legacy standards, uniformly distributed transmit symbols are employed, leading again to a shaping loss. There are different options to combat this loss, and probabilistic shaping (PS) is a promising solution for this. The readers are referred to \cite{bocherer2015bandwidth} and the references therein for a detailed study. 

PS for polar codes is recently studied in \cite{lnt_hwdu}, where a distribution matcher is used for signal shaping and a precoder is employed for systematic polar encoding. With this approach, the transmit symbols can have the optimal symbol distribution, allowing to completely compensate for the shaping loss. The receiver needs to run a multi-stage demapper/decoder and a distribution dematcher to retrieve the transmitted message. 
In this work, we propose a different approach for probabilistic shaping for polar codes and show how it can be implemented by modifying the existing 5G NR standard. Our proposal is mainly related to transmitter, and allows re-using the existing hardware for signal shaping.
We start in Sec. \ref{sec:sysmod} with the achievable rates for BICM, and show that even if signal shaping with a rough approximation of the optimal distribution is employed, a large gain compared to uniform signaling can be obtained. We then describe in Sec. \ref{sec:5gcodmod} the coding and modulation steps of 5G NR control channels with polar codes, and propose modifications in Sec. \ref{sec:modi} that allow the channel input symbols to follow an approximate Gaussian distribution. We conclude our work with simulation results and discussions in Sec. \ref{sec:numeval}.

In this paper, $\mathbf{x}$ is a vector, $\mathsf{X}$ is a random variable representing the elements in $\mathbf{x}$ and $\text{P}_{\mathsf{X}}$ describes its probability mass function. $\mathcal{D}$  denotes a set, and  $\mathbf{x}_{[\mathcal{D}]}$ is a subvector of $\mathbf{x}$ where $\mathcal{D}$ defines the indices of the elements from $\mathbf{x}$.
$\mathbb{E}(\mathsf{X})$ and $\mathbb{H}(\mathsf{X})$ denote the expected value  and the entropy of $\mathsf{X}$. We follow the notation in \cite{chan_code5G} for variable names when possible.
\begin{figure}
	\centering
	\includegraphics{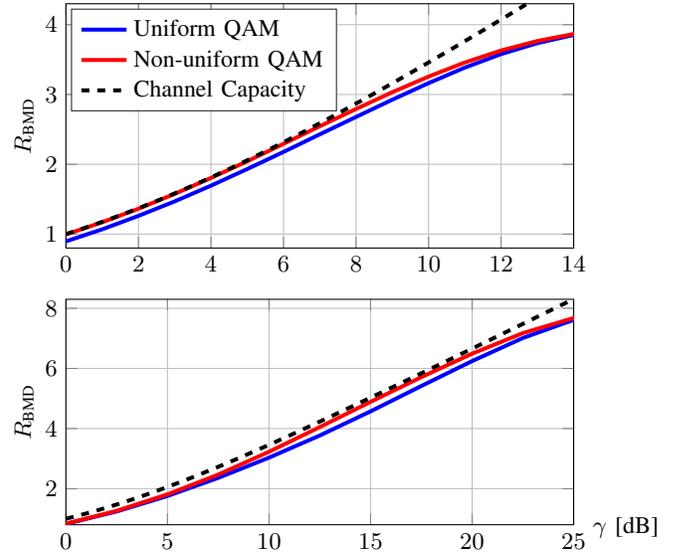}
	\caption{Achievable rates for Gray labeled 16-QAM (upper figure) and 256-QAM (lower figure) with uniform distribution and with non-uniform distribution given in (\ref{eq:pmf}) with optimized $p$.}
	\label{fig:AchRates_m4}
\end{figure}
\section{System Model and Achievable Rates}
\label{sec:sysmod}
Consider a BICM system, where the encoder generates the length-$E$ codeword $\mathbf{e}$ from the message vector $\mathbf{a}$ of length $A$. Then, an interleaver reorders $\mathbf{e}$ to the vector $\mathbf{b}$ of the same length and subsequently a symbol mapper maps $\mathbf{b}$ to $\mathbf{x}$ containing the Gray labeled $2^M$-QAM symbols with complex values consisting of real and imaginary parts from the set $\{\pm1, \pm3, ..., \pm(2^{M/2} - 1) \}$.  Here, $M$ is the number of bit-levels of the QAM symbol, e.g., 256-QAM has 8 bit-levels, and bits with the indices $\mathcal{B}_j=\{j,j+M,j+2M,...\}$ in $\mathbf{b}$ (denoted by $\mathbf{b}_{[\mathcal{B}_j]}$) define the $j$th bit-level with $j=\{1,...,M\}$. 

We consider the complex valued AWGN channel
$\mathbf{r} = \mathbf{x} + \mathbf{w}$,
where $\mathbf{w}$ is the additive noise and $\mathbf{r}$ contains the received samples after matched filtering. We define the signal-to-noise ratio as $\gamma=\mathbb{E}( \mathsf{|X|^2}  )/ \mathbb{E}(\mathsf{|W|^2})$. 
If a receiver with bit-metric decoding is considered, the (positive) rates 
\begin{align}
\label{eq:Rbicm}
\rbicm = \mathbb{H}(\mathsf{X}) - \sum_{j=1}^M \mathbb{H}(\mathsf{B}_{[\mathcal{B}_j]}|\mathsf{R})
\end{align} 
in bits/use are achievable \cite{bocherer2014achievable}. Fig. \ref{fig:AchRates_m4} shows $\rbicm$ for uniformly distributed QAM symbols for 16-QAM ($M=4$) and  256-QAM ($M=8$). For both examples, we observe a certain gap to the capacity, which is caused by two reasons. Firstly, the channel input symbols do not have the capacity achieving distribution (shaping loss), and secondly the demapper does not take the dependence between bit-levels into account (demapping loss). It was shown in \cite{bocherer2014achievable} that by using signal shaping with Maxwell-Boltzmann (MB) distribution $(\text{P}_{\mathsf{X}}(x) \sim  e^{-\nu x^2})$ for non-continuous input alphabets, this gap vanishes for a large range of rates on AWGN channels, even if independent demapping is performed. 
However, obtaining transmit symbols with an exact MB distribution in practice can be cumbersome, since this target distribution in general cannot be formulated as a product of individual bit-level distributions, and therefore symbol level shaping encoders are required. By relaxing this condition, approximations of MB distribution can be obtained with bit-level shaping encoders \cite{pikus2017bit}. In the following, we consider an approximation of the MB distribution that is generated by shaping one bit-level per complex dimension.

\begin{figure}
	\centering
	\includegraphics{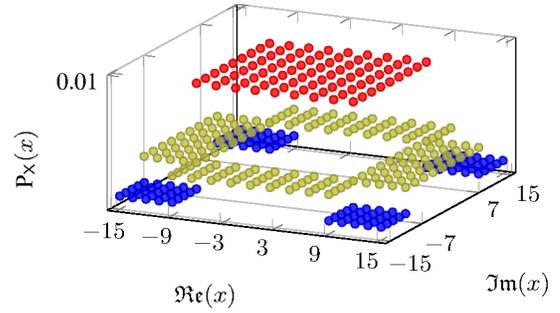}
	\caption{Illustration of $\text{P}_{\mathsf{X}}(x)$ for $M=8$ and $p=0.25$.}
	\label{fig:ConsProb}
\end{figure}

Consider a Gray symbol mapper, where the bit-level $\mti$ determines whether the magnitude of real part of the QAM symbol is larger than $2^{\frac{M}{2}-1}$ or not, i.e., the real part of the QAM symbol has an amplitude larger than $2^{\frac{M}{2}-1}$, if the bit in the  bit-level $\mti$ is $1$.
Similarly, let the bit-level $\mtq$ determine the same property for the magnitude of the imaginary part of the QAM symbol. 
If the transmitter can generate the vector $\mathbf{b}$, where the probability of bits being 1 in the subvectors $\mathbf{b}_{[\mathcal{B}_{\mti}]}$ and $\mathbf{b}_{[\mathcal{B}_{\mtq}]}$ is $p\leq0.5$, 
and the rest of the bits have equal probability of being $1$ or $0$, then the resulting QAM symbols have a piecewise constant probability distribution that can be described as
\begin{align}
\label{eq:pmf}
\text{P}_{\mathsf{X}}(x) = \left\{
\begin{array}{ll}
\dfrac{(1-p)^2}{2^{M-2}},  		&\text{if } \left\lbrace  \begin{array}{l} |\mathfrak{Re}(x)|<2^{\frac{M}{2}-1} \\ |\mathfrak{Im}(x)|<2^{\frac{M}{2}-1} \end{array} \right. \\
\dfrac{p^2}{2^{M-2}} 	 ,		&\text{if } \left\lbrace \begin{array}{l} |\mathfrak{Re}(x)|>2^{\frac{M}{2}-1} \\ |\mathfrak{Im}(x)|>2^{\frac{M}{2}-1} \end{array} \right. \\
\dfrac{p(1-p)}{2^{M-2}} ,		& \text{otherwise.}
\end{array}
\right.
\end{align}
Accordingly, a coarse approximation of MB distribution can be obtained for $p<0.5$. An example is given in Fig.~\ref{fig:ConsProb} for $p=0.25$. 
Obviously, the choice of $p$ determines the achievable rate $\rbicm$, and there is a different optimal value of $p$ that maximizes $\rbicm$ for each $\gamma$. Fig. \ref{fig:AchRates_m4} also depicts the achievable rates with (numerically) optimized $p$, and we observe that the performance gets closer to the channel capacity. Note that for $p<0.5$, the first term  in (\ref{eq:Rbicm}) gets smaller. However, the average transmit power  decreases as the symbols with high magnitudes are transmitted less frequently, which results in an overall rate increase in (\ref{eq:Rbicm}).

A key operation in obtaining the distribution in (\ref{eq:pmf}) is to generate codewords that have the certain probability distribution of bits within the codeword. In \cite{shaped_polar} it was already shown that such polar codewords can be generated by using a decoder as a precoder. 
In the following, we first describe how polar coding is employed in 5G NR control channels, and then we show modifications that allow a shaping gain for 5G polar codes by extending the ideas from \cite{shaped_polar}.

\section{Coding in 5G NR Control Channels}
\label{sec:5gcodmod}
Below we summarize coding and modulation steps for 5G NR control channels \cite{chan_code5G}. For the sake of simplicity, we consider a single transport block without code block segmentation.
\subsubsection{CRC Attachment} $L$ bits of CRC are appended to the payload vector $\mathbf{a}$, resulting in the vector $\mathbf{c}$ of length $K=A+L$. The CRC bits are used by the receiver for error detection and for selecting the correct codewords from the output of the list decoder.

\subsubsection{Polar Interleaving} The vector $\mathbf{c}$ is interleaved to  $\mathbf{c'}=\Pi_{\text{P}}(\mathbf{c})$ 
according to Table 5.3.1.1-1 in \cite{chan_code5G}, ensuring that CRC bits are distributed within the codeword allowing early termination.

\subsubsection{Polar Encoding} In this step, first a vector $\mathbf{u}$ of length $N$ is generated ($N\leq 1024$ being an integer power of two), which serves as the input to the polar transform. 
$\mathbf{c'}$ (and optionally a few additional parity check bits generated from $\mathbf{c'}$) are mapped to the reliable indices in $\mathbf{u}$ according to the polar sequence $\mathcal{Q}$ \cite[Table 5.3.1.2-1]{chan_code5G}, which is a set ordered according to the reliabilities of the polar sub-channels. The rest of $\mathbf{u}$ is filled with zeros that serve as frozen bits. Afterwards, polar transform is performed on $\mathbf{u}$ to generate the codeword
$\mathbf{d=uG}$,
where $\mathbf{G}$ is the $(\log_2 N)$-th Kronecker power of $\mathbf{G}_2=\begin{psmallmatrix}1 & 0\\ 1 & 1\end{psmallmatrix}$.

\subsubsection{Sub-block Interleaving} The sub-block interleaver reorders the vector $\mathbf{d}$ to the vector $\mathbf{y}=\Pi_{\text{SB}}(\mathbf{d})$ according to Table 5.4.1.1-1 in \cite{chan_code5G} with the aim of simple puncturing and shortening. 

\subsubsection{Bit Selection} This step generates the rate matched vector $\mathbf{e}=\mathbf{y}_{[\mathcal{E}]}$ of length $E$ where  $\mathcal{E}$ is defined as
\begin{align}
\mathcal{E} = \left\{
\begin{array}{ll}
\{ 1,\cdots,N,1,2,\cdots \},  	&\text{if } E>N  \\
\{ N-E+1,\cdots,N \},  		&\text{else if } K/E\leq7/16  \\
\{ 1,\cdots,E \},  			&\text{otherwise.}  \\
\end{array}
\right.
\end{align}
If $E$ is larger than $N$, repetition is performed. Otherwise, depending on the rate $K/E$, either the first $N-E$ bits are punctured, or the last $N-E$ bits are shortened.

\subsubsection{Code-bits Interleaving} $\mathbf{e}$ is interleaved to $\mathbf{f}=\Pi_{\text{CB}}(\mathbf{e})$ with a triangular interleaver according to \cite[Sec. 5.4.1.3]{chan_code5G}.

\subsubsection{Scrambling} Scrambling is done to by an element-wise modulo-2 sum of $\mathbf{f}$ with the user specific binary scrambling vector $\mathbf{v}$ to obtain the length-$E$ vector $\mathbf{b} = \mathbf{f}\oplus\mathbf{v}$. 
\subsubsection{QAM Mapping} $\mathbf{b}$ is mapped to channel input symbols $\mathbf{x}$ according to Gray labeling, described in \cite[Sec. 5.1]{mod5G}. Note that for this constellation $\mti=3$ and $\mtq=4$ for both 16-QAM and 256-QAM.

Depending on the scenario, these steps can be operated with different parameters, e.g., one uses only QPSK for downlink and no code-bit interleaving is used, and some additional parity check (PC) bits are appended before polar transform in uplink.
In the following, we will use these steps as a baseline and modify the model slightly to show how signal shaping can be integrated to 5G polar codes.
For simplicity, we do not consider the additional PC bits before polar transform, although our proposal can also support PC bits. We do not consider shortening and puncturing, such that $E=N$, and we will assume that 16-QAM or 256-QAM is employed, as signal shaping can only be beneficial with higher order modulation.

Note that the BICM system model described in Sec. \ref{sec:sysmod} does not contain a scrambling operation.
Let us define a length-$N$ vector $\mathbf{\bar{v}}$ satisfying
\begin{align}
\label{eq:newscrambler}
\mathbf{v} = \Pi_{\text{SB}} \Big(  \left[ \Pi_{\text{CB}}(\mathbf{\bar{v}})\right]_{[\mathcal{E}]} \Big). 
\end{align}
We can show that scrambling with $\mathbf{v}$ after code-bit interleaving is equivalent to scrambling with $\mathbf{\bar{v}}$ after polar encoding (on $\mathbf{d}$). Considering this equivalent model,  steps \textit{1)} to \textit{5)} (including scrambling with $\mathbf{\bar{v}}$) would correspond to encoding, step \textit{6)} to interleaving and step \textit{8)} to the symbol mapping in Sec. \ref{sec:sysmod}. Therefore, the achievable rates described in the previous section are also valid for this architecture.

\section{5G NR Polar Coding with Integrated Signal Shaping}
\label{sec:modi}
\begin{figure}
	\centering
	\includegraphics{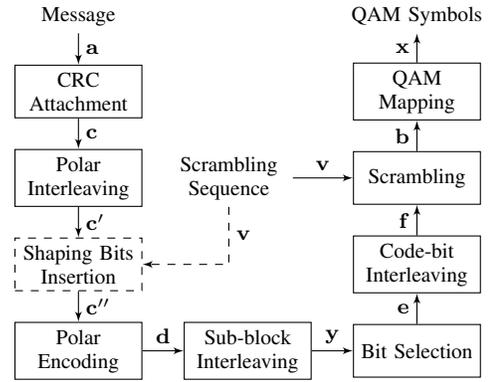}
	\caption{Block diagram describing the coding and modulation operations for 5G control channels. The dashed block performs the additional operation for signal shaping proposed in this work.}
	\label{fig:BlcDiagram}
\end{figure}

We propose two modifications to the 5G NR polar codes that allow the receiver to benefit from signal shaping gain. First, we introduce a new block called the \textit{shaping bits insertion}, which extends $\mathbf{c'}$ to $\mathbf{c''}$ by appending $S$ shaping bits $\mathbf{s}$. Those bits force the bits at the indices $\mathcal{D}=\{ N-N/(M/2)+1,...,N  \}$ of the polar codeword $\mathbf{d}$ (the last $N/(M/2)$ indices in $\mathbf{d}$) to have a certain property allowing non-uniform distribution of the QAM symbols. 
Second, the \textit{Code-bit Interleaving} step is modified such that the bits in $\mathcal{D}$ are mapped to $\mathcal{B}_{\mti}$ and $\mathcal{B}_{\mtq}$, allowing the distribution of QAM symbols given above. Fig.~\ref{fig:BlcDiagram} depicts the block diagram. 

Note that the inserted shaping bits are treated by the succeeding steps as information bits, and this operation is transparent to these steps at the transmitter, which may basically assume that the CRC appended payload has the length $A+L+S$, instead of $A+L$.

\subsection{Shaping Bits Insertion}
\label{sec:shapbit_ins}
In general, the probability of the bits being $1$ or $0$ within a binary codeword $\mathbf{d}$ is usually equal, i.e., $\text{P}_{\mathsf{D}}(1)=0.5$. In \cite{mondelli2014achieve}, it was shown how to  generate polar codewords that have a biased distribution of bits, i.e., $\text{P}_{\mathsf{D}}(1)\neq0.5$. 
The basic idea is to use some of the reliable indices of the polar transform input $\mathbf{u}$ to transmit shaping bits (in addition to information bits), which are generated depending on the information and frozen bits. The shaping bits do not carry any new information, but force $\text{P}_{\mathsf{D}}(1)$ to have a target value $p$.

In \cite{shaped_polar} it was further demonstrated that the problem of obtaining shaping bits $\mathbf{s}$ can be formulated as a polar decoding problem, i.e., a polar decoder (such as a successive cancellation list (SCL) decoder \cite{Tal15}) can be used as a precoder to generate the shaping bits from the information and frozen bits. To accomplish this, first a set $\mathcal{S}$ of reliable indices is selected. Then, a SCL decoder is used as a precoder, where the decoder treats the information and frozen bits as known (frozen) bits and seeks for bits $\mathbf{s}$ in $\mathcal{S}$ that produce a codeword with a target $\text{P}_{\mathsf{D}}(1)=p$. After obtaining the shaping bits $\mathbf{s}$, the polar transform can be performed on the vector containing frozen, information and shaping bits to generate the polar codeword having the target probability distribution.  Readers are referred to \cite{shaped_polar} for a more detailed description.

Recall that in order to obtain the probability distribution in (\ref{eq:pmf}), only the codeword bits corresponding to the bit-levels $\mti$ and $\mtq$ need to have a biased probability distribution. Similar to \cite{shaped_polar}, we exploit the idea that each polar codeword can be described as smaller codewords that are further polarized, which is depicted in Fig. \ref{fig:PolarPrecoder} with four length $N/4$ codewords and two additional polarization steps. Observe that the last small polar codeword $\mathbf{d}_{[\mathcal{D}]}'$ appears unchanged at the output as $\mathbf{d}_{[\mathcal{D}]}$. Therefore, it is enough to perform precoding only on the last small polar codeword of length $N/(M/2)$,  which can later be mapped to the correct bit-levels of the QAM symbols.

Note that the choice of $S$ and resulting $\text{P}_{\mathsf{D}}(1)=p$ are related to each other, and asymptotically  
\begin{align}
\label{eq:asym}
S=\lfloor N/(M/2)(1-\text{h}_2(p)) \rfloor
\end{align}
should hold \cite{shaped_polar}, where $\text{h}_2(.)$ is the binary entropy function. Moreover, the set $\mathcal{S}$ needs to be chosen from the reliable polar sub-channels. In this work we assume that the most reliable $S$ indices from the polar sequence $\mathcal{Q}$ in \cite{chan_code5G} define $\mathcal{S}$. We show the relation between $S$ and $p$ for finite lengths in Sec. \ref{sec:RelSP}.

Assuming the relation between $S$ and $p$ is known, we perform the following steps to obtain the shaping bits:
\subsubsection*{Step A - Construct $\mathbf{u}'$} First, we construct an auxiliary empty vector $\mathbf{u}'$ of length $N$. The most reliable $S$ indices of $\mathbf{u}'$ according $\mathcal{Q}$ are left empty. From the remaining indices, the most reliable indices are filled with $\mathbf{c}'$, and the rest are filled with zeros as frozen bits.

\subsubsection*{Step B - Obtain $\mathbf{s}$} We extract the subvector $\mathbf{\hat{u}} = \mathbf{u}_{[\mathcal{D}]}'$ of length $N/(M/2)$. We then use a polar SCL decoder of length $N/(M/2)$ where we treat the empty indices of $\mathbf{\hat{u}}$ as the set of non-frozen indices, and use the other elements in $\mathbf{\hat{u}}$ as frozen bits. We use the following LLR based decoder input
\begin{align}
\mathbf{\Lambda} = \log \left( \frac{1-p}{p} \right) (-2\cdot  \mathbf{\bar{v}}_{[\mathcal{D}]}+1),
\end{align}
where $\mathbf{\bar{v}}$ is defined in (\ref{eq:newscrambler}).
The decoder output would contain the shaping bits $\mathbf{s}$, which can be appended to $\mathbf{c'}$ to obtain $\mathbf{c}''$.

\begin{figure}
	\centering
	\includegraphics{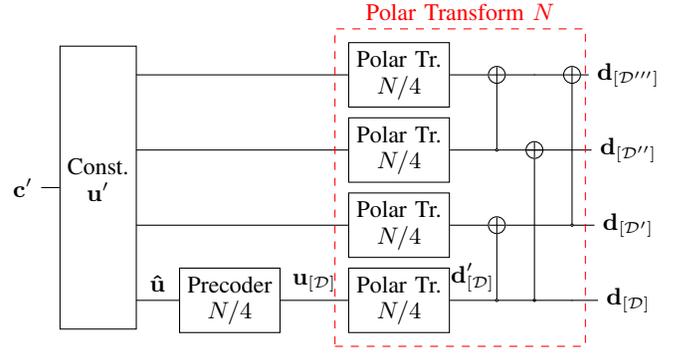}
	\caption{Polar precoding example for $M=8$, where the precoder generates the shaping bits $\mathbf{s}$ by using a polar decoder of length $N/4$, which is then used to construct $\mathbf{u}_{[\mathcal{D}]}$.}
	\label{fig:PolarPrecoder}
\end{figure}

Note that one can also fill the empty indices of $\mathbf{u}'$ with $\mathbf{s}$ to obtain $\mathbf{u}$, which can be fed to the polar transform of length $N$ to produce the codeword $\mathbf{d}$ as shown in Fig. \ref{fig:PolarPrecoder}. The generated shaping bits ensure that the bits in vector $\mathbf{d}_{[\mathcal{D}]}\oplus \mathbf{\bar{v}}_{[\mathcal{D}]}$ is $1$ with probability $p$. The effect of the scrambler is taken into the account by using $\mathbf{\bar{v}}$ to construct $\mathbf{\Lambda}$, such that the sign of the LLR values depend on the elements of $\mathbf{\bar{v}}$.

Note that the computational complexity of a length-$N$ SCL decoder with list size $L_{\text{d}}$ is given as $O(L_{\text{d}} N \log_2 N)$ \cite{Tal15}. As the precoder uses an SCL decoder of length $N/(M/2)$, its complexity is approximately $45\%$ and $20\%$ of the decoder complexity at the receiver for $N=1024$ with 16-QAM and 256-QAM, respectively, i.e. the additional complexity due to signal shaping is less than the decoding complexity at the receiver.
 
\subsection{Code-bit Interleaving}
By generating the codeword $\mathbf{d}$ as described above, bits at the indices $\mathcal{D}$ obtain the property that will cause the QAM symbols to have the desired distribution, if those bits (after interleaving and scrambling) are mapped to the bit-levels $\mti$ and $\mtq$. To guarantee correct mapping, one should also modify at least one of the interleavers. We leave the sub-block interleaver untouched, and modify the code-bit interleaver to accomplish this aim.
To make the minimum change to the existing triangular interleaver $\Pi_{\text{CB}}(.)$,  we only exchange indices in the interleaver pattern, such that $\mathbf{d}_{[\mathcal{D}]}$ is mapped to the correct bit-levels. 
We leave further optimization of the code-bit interleaver as future work. 

\subsection{Modifications at the Receiver}
The proposed signal shaping method is mainly transmitter-related, and the modifications at the receiver are small. The receiver needs to know the number of shaping bits $S$ which can be signaled without a large overhead. From $S$, the receiver can deduct $p$ (e.g. by using a look-up table) and calculate $\text{P}_{\mathsf{X}}$, which should be taken into account during demapping.
The decoding can be performed assuming $A+S$ information bits are transmitted, and the last $S$ bits can be discarded after decoding, as they do not carry any information. Optionally, the receiver can generate an estimate of $\mathbf{s}$ from the decoded information bits and cross-check its value with the received shaping bits to perform an additional error detection.

\section{Numerical Evaluation and Discussions}
\label{sec:numeval}
\subsubsection{Relation between $S$ and $p$}
\label{sec:RelSP}
In this work, we build $\mathcal{S}$ by using the most reliable $S$ indices from $\mathcal{Q}$ in \cite{chan_code5G}. To obtain $S$ for given $p$, $N$ and $M$, we propose the following steps, which can be performed offline.
\begin{itemize}
	\item Set the temporary variable $\hat{S}=1$.
	\item Randomly construct binary vectors $\mathbf{c'}$ of length $N-\hat{S}$.
	\item Perform \textit{Step A} and \textit{Step B} in Sec. \ref{sec:shapbit_ins} to obtain the shaping bits $\mathbf{s}$ and polar transform input vector $\mathbf{u}$.
	\item Obtain $\mathbf{d}=\mathbf{u}\mathbf{G}$ for each realization and calculate the average probability ($p_{\hat{s}}$) of $1$s in $\mathbf{d}_{\mathcal{D}}$.  
	\item Increase $\hat{S}$ by one and repeat the previous steps, until the maximum possible $\hat{S}$ is achieved.
	\item Set $S = \arg\min_{\hat{s}}\{|p_{\hat{s}}-p|\}$.
\end{itemize}

\begin{figure}
	\centering
	\includegraphics{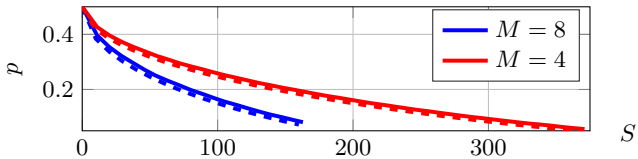}
	\caption{The relation between $S$ and $p$ obtained numerically (solid lines) and asymptotically according to (\ref{eq:asym}) (dashed lines).}
	\label{fig:shap_perf}
\end{figure}

Performing this procedure for $N=1024$ and $M=4$ and $M=8$ with a SCL decoder (list size 8), we obtain the relation show in Fig. \ref{fig:shap_perf}. We observe that by increasing the number of $S$ we obtain codewords with less ones on average. The same figure also plots the asymptotic results given in (\ref{eq:asym}). 
We observe that asymptotically one can get the same distribution by using less shaping bits. We can also see (\ref{eq:asym}) as an approximation of our numerical results. The difference between (\ref{eq:asym}) and the numerically obtained results differ on average 8 bits for $M=8$ and 14 bits for $M=4$.

\subsubsection{The Optimal Number of Shaping Bits}
In general, one can obtain the optimal value of $p$ (and hence $S$) and $A$ numerically by maximizing (\ref{eq:Rbicm}) at a given operating SNR. Here, we restrict ourselves to the polar code construction described in \cite{chan_code5G}, and obtain the optimum value of $S$ for a given $A$ with Monte-Carlo simulations.

\begin{figure}
	\centering
	\includegraphics{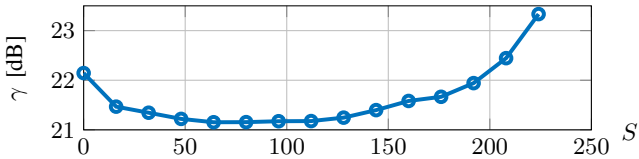}
	\caption{Performance of $A=768$ with 256-QAM on AWGN channels for different choices of $S$ in terms of required $\gamma$ to obtain a target BLER $0.001$.}
	\label{fig:k768n1024m8}
\end{figure}

As an example, we use $A = 768$ information bits, $24$ CRC bits (CRC24C from \cite{chan_code5G}) and  choose $N=E=1024$ to avoid puncturing or shortening. We use a SCL decoder with list size 8 both for precoder and for the decoder, and evaluate the block error rate (BLER) performance with different number of shaping bits $S$ for 256-QAM ($M=8$). Fig. \ref{fig:k768n1024m8} plots the required SNR to achieve a BLER of $0.001$ on AWGN channels. Note that $S=0$ corresponds to conventional BICM. We observe that by using $S=64$ shaping bits, the performance is improved by almost 1dB. The curves for other choices of $A$ and $M$ (not shown in this paper) look similar, indicating an optimal value of $S$ that is larger than $0$. 

\subsubsection{BLER Performance for the Optimal Choice of $S$}
\begin{figure}
	\centering
	\includegraphics{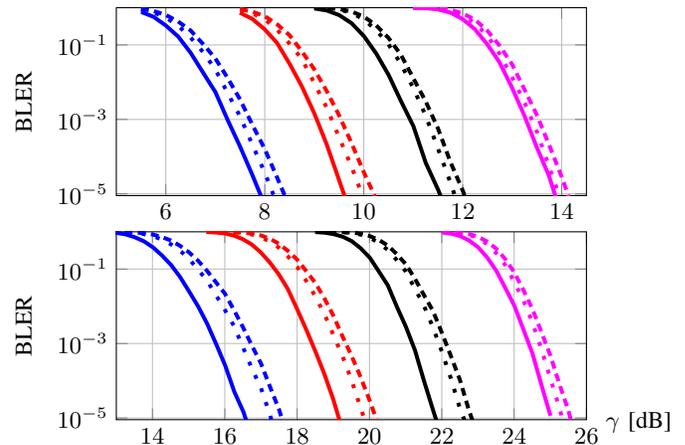}
	\caption{Upper figure: BLER performance with 16-QAM for $A = \{512, 640, 768, 896\}$ information bits, with $S = \{128, 96, 56, 16\}$ shaping bits (solid lines) and without shaping (dashed lines) with precoder and decoder list sizes 8 and $N=1024$. Dotted lines show the BLER performance without shaping with list size 16. Lower figure shows the results for 256-QAM, where the optimal number of shaping bits are $S = \{136, 120, 64, 48\}$ for the same choices of $A$ and $N$. }
	\label{fig:16QAM_bler}
\end{figure}

We use $A = \{512, 640, 768, 896\}$ information bits and find the optimal $S$ as above, leaving other parameters the same. We then evaluate the BLER performance on AWGN channels and compare it with the conventional BICM ($S=0$) for 16-QAM and and 256-QAM, as shown in Fig. \ref{fig:16QAM_bler}.  We observe $\{0.44, 0.49, 0.47, 0.25\}$dB gains compared to BICM without shaping at the target BLER of $0.001$ for 16-QAM. Note that one could expect $\{0.36, 0.45, 0.45, 0.53\}$dB gains asymptotically, considering the achievable rates in Fig. \ref{fig:AchRates_m4}. Similarly, we observe $\{0.97, 0.94, 0.93, 0.46\}$dB gains for 256-QAM. The asymptotically expected gains in this case according to Fig. \ref{fig:AchRates_m4} are $\{0.58, 0.86, 0.92, 0.82\}$dB. 
As an additional reference, we also plot the BLER performance for $S=0$ with an increased decoder list size of 16. We observe that even in this case our scheme outperforms the conventional BICM.

The presented scheme is attractive for two reasons. First, the proposed scheme performs signal shaping by using a polar decoder,  which already exists in the transmission chain of bi-directional communication systems (i.e., no new hardware is required). Second, the receiver treats the shaping bits as information bits, which can be discarded after decoding, and therefore there is no need for an additional shaping decoder (unlike other PS schemes). As the additional computation is performed at the transmitter, this method particularly suits to downlink transmissions, where the transmitter (i.e., gNB) usually have more computational power. 

\section*{Acknowledgment}
This work has been partly performed in the framework of the H2020 project ONE5G (ICT-760809) receiving funds from the EU. The views expressed in this work are those of the authors and do not necessarily represent the project view.

\bibliographystyle{IEEEtran}
\bibliography{IEEEabrv,mybibfile}

\end{document}